\documentclass[12pt, onecolumn, conference]{IEEEtran}
\IEEEoverridecommandlockouts
\usepackage{amsmath,amssymb}
\usepackage[mathscr]{eucal}
\usepackage{cite,verbatim}
\usepackage{epsfig}

\newcommand{\calX}{{\mathcal X}}
\newcommand{\calY}{{\mathcal Y}}
\newcommand{\CC}{{\mathcal C}}
\newcommand{\Cif}{\CC_{IFC}}
\DeclareMathOperator{\Exp}{{\mathbb E}} 
\newcommand{\abs}[1]{\lvert#1\rvert}

\newtheorem{thm}{Theorem}[section]

\newtheorem{cor}[thm]{Corollary}

\newtheorem{rem}{Remark}[section]

\begin{document}

\title{On the Capacity of Gaussian Weak Interference Channels with Degraded Message sets}
\author{Wei Wu, Sriram Vishwanath and Ari Arapostathis
\thanks{The authors are with Wireless Networking and Communications Group, 
Department of Electrical and Computer Engineering, The University of
Texas at Austin, Austin, TX 78712, USA (e-mail: \{wwu,sriram,ari\}@ece.utexas.edu).}
}
\maketitle

\begin{abstract}
This paper is motivated by a sensor network on a correlated field where nearby sensors share information, and can thus assist rather than interfere with one another. We consider a special class of two-user Gaussian interference channels (IFCs) where one of the two transmitters knows both the messages to be conveyed to the two receivers. Both achievability and converse arguments are provided for a channel with Gaussian inputs and Gaussian noise when the interference is weaker than the direct link (a so called weak IFC). In general, this  region serves as an outer bound on the capacity of weak IFCs with no shared knowledge between transmitters.
\end{abstract}


\section{Introduction}

An interference channel (IFC), characterized by the channel $p(y_1,y_2|x_1,x_2)$, is one of the basic building blocks of many networks and is thus considered a fundamental problem in multi-user information theory. However, the capacity of this channel remains an open problem, with some special cases such as the strong-interference case being characterized. One of the fundamental difficulties faced while attacking the IFC capacity problem is that, unlike the broadcast channel, no transmit-side cooperation is possible.

	In this paper, we consider a two-user Gaussian IFC where we allow limited cooperation between the transmitters by means of permitting one of the transmitters to possess the message of the other. Such an cooperative IFC is of interest of its own merit, as it represents interference channels resulting from systems such as sensor networks, where the two transmitters gather correlated information. It is also interesting as its capacity region is an outer bound on that of the non-cooperative IFC.

\begin{figure}[hbtp]
\centering
\includegraphics[width=3in]{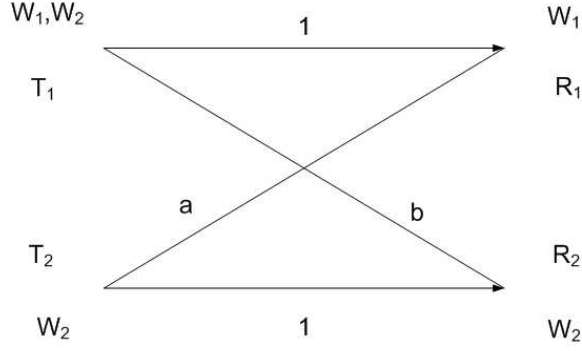}
\caption{The interference channel with partial transmission cooperation}
\label{fig:coop_if}
\end{figure}

This Gaussian IFC is shown schematically in Figure ~\ref{fig:coop_if}. In this system $T_1$ and $T_2$ represent two transmitters (sensors). One of the transmitters ($T_1$) is in possession of both messages $w_1$ intended for Receiver 1 and $w_2$ intended for Receiver 2, while the other transmitter only possesses $w_2$.
 The additional information ($w_2$) at $T_1$ can help improve the transmission at $T_1$ in two ways: the interference seen  from $T_2$ at $R_2$ can be mitigated by a suitable precoding operation at $T_1$ (dirty paper coding); and $T_1$ can aid $T_2$ in the transmission of $w_2$ (cooperative transmission).

Let $\Cif^{T_{i}}$ represent the capacity of the cooperative IFC whee $T_i, i \in 1,2$ possesses both messages. Then the capacity of the non-cooperative IFC $\Cif$ can be outer bounded by
\begin{equation} \label{eq:T1_T2}
\Cif \subset \Cif^{T_{1}} \bigcap \Cif^{T_{2}}
\end{equation}

A memoryless IFC is formally defined by the alphabets $\calX_1$,$\calX_2$, $\calY_1$, $\calY_2$, and the conditional probability distribution $P(y_1, y_2| x_1, x_2)$ where $x_t \in \calX_t$ and $y_t \in \calY_t$, $t=1,2$.
The channel is time invariant and memoryless in the sense that
\begin{equation}
P(y_{1,n},y_{2,n}|x_1^n, x_2^n, y_1^{n-1}, y_2^{n-1})=P_{Y_1 Y_2|X_1 X_2}(y_{1,n},y_{2,n}|x_{1,n},x_{2,n})\,,
\end{equation}
where $x_t^n$ denotes the history of the sequence $x_{t,k}$, $k$ runs up to $n$.  The capacity region is defined to be the closure of the set of rate pairs $(R_1,R_2)$ for which the receivers can decode their respective messages at an arbitrarily small error probability. 

Our interest in this domain is the Gaussian IFC in which the alphabets of inputs and outputs are real numbers and the outputs are linear combinations of input signals and white Gaussian noise. 
The  Gaussian IFC is defined as follows,
\begin{equation}
\begin{split}
Y_1 &=X_1+a X_2 + Z_1 \\
Y_2 &=b X_1 + X_2 + Z_2
\end{split}
\end{equation}
where $a$ and $b$ are real numbers and $Z_1$ and $Z_2$ are independent, zero-mean, unit-variance Gaussian random variables.
Furthermore, the transmitters are subject to average power constraints:
\begin{equation}
\lim_{N \to \infty} \frac{1}{N}\sum_{n=1}^{N} \Exp[X_{t n}^2] \leq P_t\,, \quad t=1,2\,.
\end{equation}

The capacity  region of the non-cooperative Gaussian IFC is currently known in single-letter form only when $a=b=0$ (a trivial case) or if $a^2 \geq 1$ and $b^2 \geq 1$ (the so called \emph{strong interference} case). 
The capacity of IFC for strong interference is the set of $(R_1, R_2)$ satisfying \cite{Sato:IT81StrongIF,Han:IT81StrongIF}
\begin{align}
0 &\leq R_1 \leq \frac{1}{2}\log(1+P_1)\label{eq:IFC_R1}\\
0 &\leq R_2 \leq \frac{1}{2}\log(1+P_2)\label{eq:IFC_R2}\\
0 &\leq R_1+R_2 \leq \frac{1}{2}\log(P_1+ a^2 P_2+1)\\
0 &\leq R_1+R_2 \leq \frac{1}{2}\log(b^2 P_1+P_2+1)\,.
\end{align}  
In the region that $0 \leq a^2 \leq 1$ or $0 \leq b^2 \leq 1$, there are no capacity results on IFC, but a set of achievable rate regions \cite{Han:IT81StrongIF, Costa:IT85IFC, Sason:IT04IFC} and capacity outer bounds \cite{Sato:IT77TwoUser, Carleial:IT83outbound, Costa:IT85IFC, Kramer:IT04outbound}.
The most recent outer bound discovered by Kramer in \cite{Kramer:IT04outbound} is $(R_1, R_2)$ satisfying \eqref{eq:IFC_R1}, \eqref{eq:IFC_R2}, and 
\begin{align}
R_1+R_2 \leq \frac{1}{2}\log \Bigl[(P_1+a^2 P_2+1)\bigl(\frac{P_2+1}{\min(a^2,1) P_2+1} \bigr) \Bigr] \label{eq:sum_T2}\\
R_1+R_2 \leq \frac{1}{2}\log \Bigl[(P_2+b^2 P_1+1)\bigl(\frac{P_1+1}{\min(b^2,1) P_1+1} \bigr) \Bigr] \label{eq:sum_T1}\,.
\end{align}

In this paper, we characterize the capacity region of Gaussian IFCs where one transmitter ($T_1$) knows both the messages {\it and} when $b^2 \le 1$. We call this category of Gaussian IFCs the weak IFCs with degraded message sets. 
To our knowledge, this is the first capacity result for Gaussian interference channels under weak interference.

\section{The capacity region of Gaussian IFC with Gaussian inputs and partial transmitter cooperation}
\begin{figure}[hbtp]
\centering
\includegraphics[width=4in]{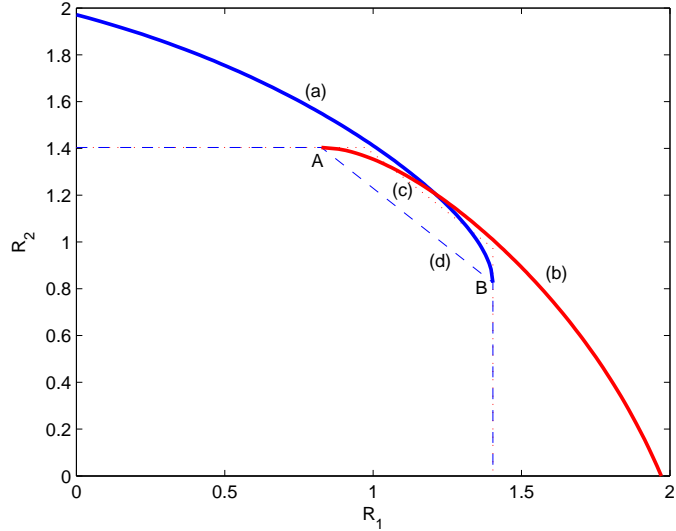}
\caption{The outer bounds of two-user Gaussian interference channel with $P_1=P_2=6$, $a^2=b^2=0.3$: (a) the capacity region $\Cif^{T_1}$; (b) the capacity region $\Cif^{T_2}$; (c) Carleial's outer bound in \cite{Carleial:IT83outbound}; (d) Kramer's outer bound in \cite{Kramer:IT04outbound} (theorem 1).}
\label{fig:ifc_outbound}
\end{figure}

\begin{thm}\label{thm:T1}
The capacity region of the Gaussian IFC with Gaussian inputs and transmitter $T_{1}$ knowing both messages, $\Cif^{T_1}$, when $\abs{b} \leq 1$, is the convex hull of the closure of all $(R_1, R_2)$ that satisfy
\begin{eqnarray}
R_1 & \leq & I(X_1; Y_1 | X_2, U) = \frac{1}{2}\log\left(1+ \alpha P_1 \right)\label{eq:cap_R1}\\
R_2 & \leq & I(U, X_2; Y_2)= \frac{1}{2} \log\left(1 + \frac{h \Sigma h^t}{1+ b^2 \alpha P_1}\right) \nonumber\\
 & =& \frac{1}{2} \log\left(\frac{1+P_1 b^2+2\abs{b}\sqrt{(1-\alpha)P_1P_2}+P_{2}}{1+ b^2 \alpha P_1} \right)  \label{eq:cap_R2}
\end{eqnarray} 
 Here $h$ is the vector $[b ~1]$, and $\Sigma$ is a $2 \times 2$ covariance with diagonal elements equaling $(1-\alpha)P_1$ and $P_2$ respectively.
\end{thm}
\begin{proof}

Achievability: The achievability of this rate utilizes the now famous dirty-paper coding strategy. First, we generate a codebook of $2^{nR_2}$ codewords according to ${\mathcal{N}}(0,\Sigma)$, where $\Sigma$ is the covariance between transmitter 1 and 2. Transmitter 1 devotes $(1- \alpha)$ fraction of its power $P_1$ to the transmission of $W_2$, while Transmitter 2 devotes its entire power $P_2$ to this effort. This leads to a covariance of the form

\begin{equation}
\Sigma = \left[\begin{array}{cc} (1-\alpha)P_1 & \gamma \\ \gamma & P_2 \end{array}\right]
\end{equation}
 
The effective interference seen by Receiver 1 is a combination of the signals communicated from both Transmitters 1 and 2. Since Transmitter 1 knows the exact realization of the message $w_2 \in W_2$, it has non-causal side information on the interference and can completely cancel it out, achieving a rate
\begin{equation}
R_1 = \frac{1}{2} \log\left(1+ \alpha P_1 \right)
\end{equation}
using a Gaussian codebook with codewords that are correlated with the interference. At Receiver 2, this Gaussian codebook for $W_1$ is perceived as additive interference, hence achieving a rate:

\begin{equation}
R_2= \frac{1}{2} \log\left(1 + \frac{h \Sigma h^t}{1+ b^2 \alpha P_1}\right)
\end{equation}
Maximize $R_2$ over $\abs{\gamma}^2 \leq (1-\alpha)P_1 P_2$ (such that $\Sigma$ is positive definite), it is not difficult to shown $R_2$ obtain the maximum when $\gamma = \sqrt{(1-\alpha)P_1 P_2}$ and \eqref{eq:cap_R2} can be achieved. 

\bigskip
Converse:
The central feature here is identifying the correct auxiliary random variable. We first determine an outer bound for the DMC case, and then replace the expressions obtained with Gaussian inputs to obtain the result.
\begin{subequations}
\begin{align}
n R_1 &= H(W_1) \leq I(W_1; Y_1^n) \leq I(W_1; Y_1^n | W_2, X_2^n) \label{eq:R1_a}\\
& \leq  \sum_{i=1}^n I(W_1; Y_{1,i}|W_2, Y_1^{i-1},X_2^{i-1}, X_{2,i}, X_{2,i+1}^n)\label{eq:R1_b}
\end{align}
The inequalities \eqref{eq:R1_a} and \eqref{eq:R1_b} is due to the reason that the mutual information will be increases by adding conditionals. 
Define $U_i=(W_2, Y_1^{i-1}, X_2^{i-1})$,
\begin{align}
n R_1 & \leq \sum_{i=1}^n I(W_1; Y_{1, i}|U_i, X_{2,i}, X_{2,i+1}^n) \\
& = \sum_{i=1}^n H(Y_{1,i}|U_i, X_{2,i}, X_{2,i+1}^n)- \sum_{i=1}^n H(Y_{1,i}| U_i, X_{2,i}, X_{2,i+1}^n, W_1)  \label{eq:R1_d}\\
& \leq \sum_{i=1}^n H(Y_{1,i}|U_i, X_{2,i})-\sum_{i=1}^n H(Y_{1,i}|U_i,X_{2,i},X_{2,i+1}^n,W_1,X_{1,i}) \label{eq:R1_e} \\
& =\sum_{i=1}^n H(Y_{1,i}|U_i,X_{2,i})-\sum_{i=1} H(Y_{1,i}|U_i,X_{1,i},X_{2,i}) \label{eq:R1_f} \\
& = \sum_{i=1}^n I(X_{1,i}; Y_{1,i}|U_i,X_{2,i}) \label{eq:R1_g}
\end{align}
\end{subequations}
\eqref{eq:R1_e} is because the entropy will increase when dropping some conditionals and it will decrease by adding more conditions.
The equality \eqref{eq:R1_f} is true because $W_1-(X_{1,i}, X_{2,i})-Y_{1,i}$ forms a Markov chain and $Y_{1,i}$ does not depend on $X_{2,i+1}^n$, the information to be transmitted in the future. 
 
Now consider $R_2$,
\begin{subequations}
\begin{align}
n R_2 &=H(W_2) \leq I(W_2; Y_2^n) = \sum_{i=1}^n I(W_2; Y_{2,i}|Y_2^{i-1}) \\
& \leq \sum_{i=1}^n H(Y_{2,i})-\sum_{i=1}^n H(Y_{2,i}|Y_2^{i-1},W_2, X_{2,i}, X_2^{i-1}) \label{eq:R2_mid1}
\end{align}
\end{subequations}
Define 
\begin{align}
Y'_{2,i}&=b X_{1,i}+Z_2 \,, \\
Y'_{1,i}&=X_{1,i}+Z_1\,.
\end{align}
By $b \leq 1$, we can see $Y'_{2,i}$ is a stochastically degraded version of $Y'_{1,i}$.
Thus
\begin{subequations}
\begin{align}
H(Y_{2,i}|Y_2^{i-1},W_2, X_2^i)& = H(Y'_{2,i}|Y_2^{i-1},W_2, X_{2,i}, X_2^{i-1}) \label{eq:R2_a}\\
&= H(Y'_{2,i}|Y_2^{' i-1},W_2, X_{2,i}, X_2^{i-1}) \label{eq:R2_b} \\
& \geq H(Y'_{2,i}|Y_1^{' i-1},W_2, X_{2,i}, X_2^{i-1}) \label{eq:R2_c}\\
& =H(Y'_{2,i}|Y_1^{i-1},W_2, X_{2,i}, X_2^{i-1}) \label{eq:R2_d} \\
& =H(Y_{2,i}|Y_1^{i-1},W_2, X_2^{i-1}, X_{2,i}) \label{eq:R2_e}\\
& =H(Y_{2,i}|U_i,X_{2,i}) \label{eq:R2_f}
\end{align}
\end{subequations}
Combine \eqref{eq:R2_mid1} and \eqref{eq:R2_f}, we get the results. To show that i.i.d coding is optimal, we formulate the optimization problem that characterizes the boundary 
\begin{equation}
\sup_{p(u^n,x_2^n,x_1^n)} R_1 + \mu R_2
\end{equation}
Note that
\begin{align*}
 \sup_{p(u^n,x_2^n,x_1^n)} n (R_1 + \mu R_2)
&\le \sup_{p(u^n,x_2^n,x_1^n)}\sum_{i=1}^n I(X_{1,i}; Y_{1,i}|U_i,X_{2,i})   +\mu  \sum_{i=1}^n I(U_i, X_{2,i}; Y_{2,i})\\
&\le \sum_{i=1}^n \sup_{p(u^n,x_2^n,x_1^n)} I(X_{1,i}; Y_{1,i}|U_i,X_{2,i})   +\mu  I(U_i, X_{2,i}; Y_{2,i})\\
&\le \sum_{i=1}^n \sup_{p(u_i,x_{2i},x_{1i})} I(X_{1,i}; Y_{1,i}|U_i,X_{2,i})   +\mu  I(U_i, X_{2,i}; Y_{2,i})\\
\end{align*}
Substituting Gaussians for the inputs, we get the result. Note that if Gaussian inputs are optimal, proving optimality is not trivial and requires arguments similar to the proof in the degraded Gaussian MIMO (multiple input multiple output) broadcast channel case \cite{Shamai:IT_BCCapacity}. For example, defining $V$ to be the vector $U,X_2$, we find that the outer bound is structured as $R_1 \le I(X_1;Y_1|V)$ and $R_2 \le I(V;Y_2)$, exact analog of the outer bound on degraded MIMO broadcast channels. 
\end{proof}
\bigskip
By swapping the parameters for $T_1$, $T_2$, the capacity region $\Cif^{T_2}$ can be obtained
\begin{cor}\label{cor:T2}
The capacity region of the Gaussian IFC with Gaussian inputs and transmitter $T_{2}$ knowing both messages, $\Cif^{T_1}$, when $\abs{a} \leq 1$, is
\begin{equation} \label{eq:cap_T2}
\Cif^{T_{2}}=\left\{
(R_{1},R_{2}): 
\begin{array}{ccc}
R_1 & \leq &\frac{1}{2} \log\Bigl(\frac{1+P_1 +2\abs{a}\sqrt{(1-\beta) P_1P_2}+a^{2}P_{2}}{1+ a^2 \beta P_2} \Bigr)  \\
R_2 & \leq &\frac{1}{2}\log\left(1+ \beta P_2 \right) 
\end{array}
\,, 0 \leq \beta \leq 1
\right\} \;.
\end{equation} 
\end{cor}
\bigskip
\begin{rem}
$\alpha=0$ in Theorem~\ref{thm:T1} corresponds to the full cooperation of $T_1$ and $T_2$ to transmit $W_2$ and the capacity \eqref{eq:cap_R2} becomes into the capacity of a MISO (multiple input single output) channel. 
\end{rem}
\medskip
\begin{rem}\label{rem:outbound}
As $\alpha=1$, the capacity \eqref{eq:cap_R1} \eqref{eq:cap_R2} given in in Theorem~\ref{thm:T1} is 
\begin{equation}
\begin{split}
C_1^{T_1}&=\frac{1}{2}\log(1+ P_1 ) \\
C_2^{T_2}&= \frac{1}{2} \log\bigl(\frac{1+P_1 b^2+ P_{2}}{1+ b^2 P_1}\bigr)\,,
\end{split}
\end{equation}
which is exactly the extreme point corresponding to \eqref{eq:IFC_R1} and \eqref{eq:sum_T1} in Kramer's outer bounds \cite{Kramer:IT04outbound}.
On the other hand, the capacity point in Corollary~\ref{cor:T2} when $\beta=1$ corresponds to the extreme point determined by \eqref{eq:IFC_R2} and \eqref{eq:sum_T2}. 
Note the outer bounds \eqref{eq:sum_T1} \eqref{eq:sum_T2} are proved based on a genie-aided argument,or ``receiver genie-aided'' approach, namely, some genie providing additional channel output to one of receivers.
In contrast, our approach here can be viewed as ``transmitter genie-aided'' in the sense that some genie gives additional information to one of the transmitters.
We comment that the outer bound by \eqref{eq:T1_T2} is not as good as the one obtained in \cite{Kramer:IT04outbound}.
However, the bound in \eqref{eq:T1_T2} might be possibly improved when one transmitter only knows some function of another message instead of the full message, i.e., in Figure~\ref{fig:coop_if} transmitter $T_1$ only know $g(W_2)$ instead of knowing $W_2$, where $g(\cdot)$ is some function of $W_2$.
\end{rem}
\bigskip
In Figure~\ref{fig:ifc_outbound}, we compare some known outer bounds for two-user Gaussian IFC for $P_1=P_2=6$ and $a^2=b^2=0.3$. The rate units are bits per channel use.
As we have discussed in remark~\ref{rem:outbound}, Kramer's outer bound in \cite{Kramer:IT04outbound} meets our outer bound of \eqref{eq:T1_T2} at point $A$, $B$, and performs better than ours elsewhere. 
Our outer bound in \eqref{eq:T1_T2} gives a better bound than Carleial's when $\alpha$ and $\beta$ in \eqref{eq:cap_R1} \eqref{eq:cap_R2} \eqref{eq:cap_T2} is close to 1. 

\section{Conclusions and Future Work}
In this paper, we studied the capacity region of a special class of two-user Gaussian interference channel (IFC) with degraded message sets, in which one transmitter knows both messages. This region is an outer bound on the capacity region of the non-cooperative Gaussian IFC and is of interest of its own merit. However, it does not improve upon known bounds in \cite{Kramer:IT04outbound}, which is based on introducing receiver cooperation. Possible extensions to this work include: (i) Lossy functions of the message ($w_2$) made available to $T_1$ rather than the message itself. This might possibly yield a better outer bound for the non-cooperative Gaussian IFC. 
(ii) Generalizing this approach to  Gaussian IFC with more than two users in the system.
\bibliographystyle{IEEEtran}
\bibliography{ifc_nfb}
\end{document}